\documentclass{pasa}%

\usepackage{graphicx}
\usepackage[version=3]{mhchem}
\usepackage[T1]{fontenc} 

\usepackage{isotope}
\usepackage{xspace}
\usepackage{multirow}
\usepackage{hyperref}
\usepackage{comment}

\newcommand{\ro}[1]{\ensuremath{\textrm{#1}}}

\newcommand{\dd}{\ensuremath{\ro{d} }}

\newcommand{\HeF}{\isotope[4]{He}\xspace}
\newcommand{\LiF}{\isotope[5]{Li}\xspace}

\newcommand{\BeE}{\isotope[8]{Be}\xspace} 
\newcommand{\BeN}{\isotope[9]{Be}\xspace} 
\newcommand{\Oxs}{\isotope[16]{O}\xspace} 
\newcommand{\Fe}{\isotope[56]{Fe}\xspace}

\newcommand{\sv}{\ensuremath{~\langle \sigma v \rangle}\xspace}
\newcommand{\ten}[1]{\ensuremath{\times 10^{#1}}}

\title[Entropy at the Start of the Universe]{Under an Iron Sky: \\On the Entropy at the Start of the Universe}

\author[Barnes and Lewis]{Luke A. Barnes$^1$\thanks{l.barnes@westernsydney.edu.au} ~and Geraint F. Lewis$^{2}$
\affil{$^1$School of Science, Western Sydney University, Locked Bag 1797, Penrith South, NSW 2751, Australia}%
\affil{$^2$Sydney Institute for Astronomy, School of Physics A28, The University of Sydney,  NSW, 2006, Australia}
}%

\jid{PASA}
\doi{10.1017/pas.\the\year.xxx}
\jyear{\the\year}

\usepackage{aas_macros}
\usepackage{hyperref} 
\hypersetup{colorlinks,citecolor=blue,linkcolor=blue,urlcolor=blue}

\hypersetup{draft}

\begin{document}

\begin{frontmatter}
\maketitle

\begin{abstract}
Curiously, our Universe was born in a low entropy state, with abundant free energy to power stars and life. The form that this free energy takes is usually thought to be gravitational: the Universe is almost perfectly smooth, and so can produce sources of energy as matter collapses under gravity. It has recently been argued that a more important source of low-entropy energy is nuclear: the Universe expands too fast to remain in nuclear statistical equilibrium (NSE), effectively shutting off nucleosynthesis in the first few minutes, providing leftover hydrogen as fuel for stars. Here, we fill in the astrophysical details of this scenario, and seek the conditions under which a Universe will emerge from early nucleosynthesis as almost-purely iron. In so doing, we identify a hitherto-overlooked character in the story of the origin of the second law: matter-antimatter asymmetry.
\end{abstract}

\end{frontmatter}

\section{Introduction}
\label{sec:intro}
The early Universe was relatively simple. A few minutes after the Big Bang, the Universe was composed of electromagnetic radiation, dark matter, electrons, and baryons, mainly in the form of hydrogen and helium \citep{Mathews_2017}. Dark energy was present, lurking in the background at a level that would be irrelevant for several billion years \citep{2014EPJC...74.3160V}.  The Universe's various forms of energy were (almost) uniformly distributed through space.

Slight inhomogeneities, perhaps seeded by inflation, were growing under the attractive pull of gravity, with dark matter and gas pooling into the sites that would become present day galaxies \citep{1978MNRAS.183..341W,Springel_2006}. As the density of gas increased, so did its ability to cool, fragment and collapse. Within this compressed gas, the first proto-stars formed as nuclear reactions ignited inside their cores. These stars ended their lives in supernovae explosions, polluting the interstellar medium with heavier elements. This enriched medium formed into future generations of stars \citep{Aguirre_2017}, which soon reionized the Universe \citep{1965ApJ...142.1633G}.

The formation and redistribution of heavier elements was essential for the emergence of life in the Universe. Stars form the raw materials for planets and creatures \citep{Armitage2018}. Stars also bathe planets with low entropy radiation, the ultimate power source for life \citep{alma991003332799705106}. Stars in our Universe emit photons whose energy is roughly the same as the typical energy of chemical bonds, making processes such as photo-synthesis possible. This fact traces back to a remarkable coincidence between fundamental constants: $m_p^3 \approx \alpha^6 ~ m_e^2 ~ m_{Pl}$, where $m_p$, $m_e$ and $m_{Pl}$ are the proton, electron and Planck masses respectively, and $\alpha$ is the fine-structure constant \citep{1974IAUS...63..291C}.

This {\it stelliferous} period of the life in the Universe, though it may last for trillions of years, is fleeting compared to the indefinite amount of time ahead, where energy will become sparse and the cosmos will tend to a state of maximum entropy \citep{2012coup.book...71A}. As with all isolated physical systems, the Universe seems headed for a state in which all its energy is trapped in useless forms, such as inside black holes or evenly dispersed in low-energy radiation.

Statistical mechanics explains this tendency --- the second law of thermodynamics -- roughly as follows. A complete description of every detail of a physical system is called a \emph{microstate}. As well as the actual microstate of a system at a particular time, we also consider the space of all possible microstates of the system (\emph{phase space}). The laws of nature describe which future microstate a given microstate will evolve into after a certain amount of time. However, for many systems of interest, this will involve an infeasibly large amount of detail. More practically, we use a statistical approach: armed with a method --- known as \emph{coarse graining} --- that calculates macroscopic quantities (e.g. temperature) from microscopic ones (e.g. the position and momentum of every particle) and a probability distribution over microstates, we can seek to derive the assured results of classical thermodynamics. We find that the space of microstates is typically dominated by equilibrium states of maximum entropy, that is, maximum entropy states are the most probable states. Thus, a system that has had sufficient time to explore its possible states is very likely to be found in a maximum entropy state. (Consult your local statistical mechanic for a more nuanced account).

This raises a puzzle: we have explained why low entropy states tend to evolve toward high entropy states, but why are there \emph{any} low entropy states? If low entropy states are improbable, why are any observed at all? This puzzle becomes all the more acute as we rewind the second law back to the beginning of the Universe: why wasn't the Universe born in a maximum entropy state? There are, after all, plenty of them! The need for a low-entropy boundary condition in the past to explain the second law of thermodynamics is known as the {\it Past Hypothesis}. We will not resolve these questions here; the interested reader is directed to \citet{Penrose1979,Huw1996,Albert2000,2004hep.th...10270C,2005gr.qc.....7094W,Earman2006a,Frigg2010,Wallace2011a,Winsberg2012,Goldstein2016a} and beyond. 

Those who subscribe to a low-entropy beginning have sought to identify the form of free energy that is available in the early Universe, that is, the aspect of the arrangement of the early Universe that supplies its low entropy. This presents another puzzle: the early Universe was a homogeneous plasma with uniform temperature, and to our thermodynamic intuition, trained on boxes of classical ideal gas, this may appear to be a state of maximum entropy.

The standard solution to this puzzle was presented in \citet{Penrose1979,1989NYASA.571..249P}, and popularised in his book {\it The Emperor's New Mind} \citep{1990enmc.book.....P}. Penrose identifies the crucial role of gravity: the highest-entropy arrangement of a box of matter is that in which all the matter has collapsed into a black hole. A high-entropy big bang would be a expanding Universe that is born full of black holes. By contrast, in our Universe, while the initial homogeneity will not allow for thermal energy to be extracted by heat flowing from hot to cold regions, gravity will cause the matter can collapse into clumps, transforming potential energy into kinetic energy. By comparing our observable Universe with a black hole of the same mass, Penrose calculates the entropy of our Universe relative to its maximum, which in turn implies that the fraction of phase space that is at least as low entropy as our Universe is extremely small, one part in $10^{10^{123}}$. 

However, \citet{2010BJPS...61..513W} and \citet{2019Entrp..21..466R} have argued that most important source of low entropy is not gravitational but nuclear. Our Universe at early times is in Nuclear Statistical Equilibrium (NSE). At high temperatures, NSE favours small nuclear species: protons and neutrons. As the temperature and density of the Universe decrease via expansion, this equilibrium is maintained as long as the proton $\leftrightarrow$ neutron reaction rate is fast compared to the expansion of the Universe. However, at about 1 second ($T \sim 0.8 MeV$) the reaction rate has slowed relative to the expansion such that the proton to neutron ratio freezes out. After this, nucleosynthesis begins in earnest about at $\sim 100$ seconds. By $\sim 1000$ seconds, the Universe has expanded and cooled to the point that nucleosynthesis is over, leaving the baryonic component in the form of hydrogen and helium, with a trace of heavier elements.

Our thermodynamic, second law arrow of time is driven largely by low-entropy (hotter than the CMB) radiation from the Sun. While the Sun was \emph{ignited} by gravitational collapse over its first $\sim$ million years as a protostar, its energy output for the last 4.5 billion years has been powered by the fusion of hydrogen to helium. However, as noted by \citet{2019Entrp..21..466R} and as we will show in later section, if the Universe had maintained NSE through just one more decade in temperature (down to $10^9~$K, instead of $10^{10}~$K as in our Universe), with the nuclear reaction rates remaining fast compared to cosmic expansion, all protons and neutrons would have been bound into heavy nuclei. The Universe would be a homogeneous plasma of \Fe. ``Stars'' formed from such strongly-bound nuclei would be unable to undergo fusion (unless their initial collapse was so violent that \Fe\  was disintegrated into smaller elements). Furthermore, with a periodic table consisting of a single element, the chemical complexity required for life would be unavailable. \citet{2019Entrp..21..466R} notes that our Universe, by failing to maintain NSE, exits nucleosynthesis in a {\it metastable state}, stranded in a low entropy state until gravitational collapse greatly accelerates the nuclear reaction chain toward the NSE state of a low temperature plasma: iron.

\citet{2019Entrp..21..466R} characterises the situation by saying that, 
\begin{quote}
``the dominant source of low-entropy that feeds the observed irreversible behaviour of the universe is a single degree of freedom, the scale factor, which is (in a precise technical sense specified below) out of equilibrium.'' 
\end{quote}
This description is correct, but it doesn't answer the question: why is the scale factor of our universe out of equilibrium? The normalisation of the scale factor is arbitrary; only the evolution of its relative value has physical meaning. The evolution of the scale factor depends on the fundamental constants of cosmology. Which of these, in concert with the physics of nuclear reactions, determines that our Universe begins out of nuclear equilibrium?

In this paper, we fill in some of the astrophysical details of the argument of \citet{2010BJPS...61..513W} and \citet{2019Entrp..21..466R}, and in so doing identify the crucial role played by an as-yet overlooked character: matter-antimatter asymmetry. The scale factor of our universe is out of equilibrium (in the sense that Rovelli explains) in part because the matter-antimatter asymmetry in our universe is small. Universes with larger matter-antimatter asymmetry burn more of their nucleons to heavy elements in their early stages.

The layout of this paper is as follows. In Section~\ref{sec:nucleo}, we revisit nucleosynthesis in our Universe, reviewing the impact of reaction rates and expansion histories. In Section~\ref{sec:nse}, we consider nuclear statistical equilibrium and its perturbation by an expanding Universe. In Section \ref{sec:ironUniverse}, we explain how to make an almost-iron Universe, and the implications for the fundamental constants of cosmology. Finally, in Section \ref{sec:lateentropy} we discuss the implications for the initial entropy of the Universe and the second law.

\section{Big Bang Nucleosynthesis}
\label{sec:nucleo}
In this section, we will review the relationship between cosmic expansion and nuclear reaction rates in the early Universe. Numerical codes such as those of \citet{2018arXiv180611095A} and \citet{2016MNRAS.460..291L} that follow primordial nucleosythesis consider only the lightest few elements in the periodic table, up to oxygen. Hence, they cannot follow the equilibrium state of the Universe all the way to iron. Thus, we will focus in this section on the equilibrium between protons and neutrons for a family of cosmic expansion histories, expanding to consider NSE in more detail in later sections.

We follow the account of nucleosynthesis in \citet{2004IJTP...43..669M}. The abundance of free neutrons is,
\begin{equation}
    X_n = \frac{n_n}{n_n + n_p} ~,
\end{equation}
where $n_n$ $(n_p)$ is the number density of free neutrons (protons). In NSE, the ratio of free neutrons to free protons is given by,
\begin{equation} \label{eq:nnnp}
    \frac{n_n}{n_p} = e^{-Q_{np}/T}
\end{equation}
where $T$ is the temperature and $Q_{np} = 1.293$ MeV is the mass difference between the neutron and proton. As shown by \citet{2004IJTP...43..669M}, the relative number density of neutrons is given by an asymptotic series of the form,
\begin{equation}
    X_n = X_n^{eq} 
    \left( 
    1 - \frac{1}{\lambda_{np}} \left( 1 + e^{-\frac{Q_{np}}{T}} \right)^{-1} \frac{{\dot{X}}_n^{eq}}{X_n^{eq}} + ...
    \right) ~.
    \label{rate}
\end{equation}
Hereafter, we will ignore the small corrections represented by the ellipses. In this equation, $\lambda_{np}$ is the neutron to proton interaction rate, given by,
\begin{equation}
    \lambda_{np} = \frac{255}{\tau_n x^5} \left( x^2 + 6x + 12\right)
\end{equation}
where $\tau_n \approx 886$ seconds is the neutron lifetime and $x = \frac{Q_{np}}{T}$ \citep{1989RvMP...61...25B}. If the rate of reaction $\lambda_{np}$ is high compared to the rate at which the expansion of the Universe perturbs the abundance away from equilibrium ($\sim \dot{X}_n^{eq} / X_n^{eq}$), then the system remains close to its equilibrium state, $X_n \approx X_n^{eq}$.

\begin{figure}[tbp]
\centering 
\includegraphics[width=.5\textwidth]{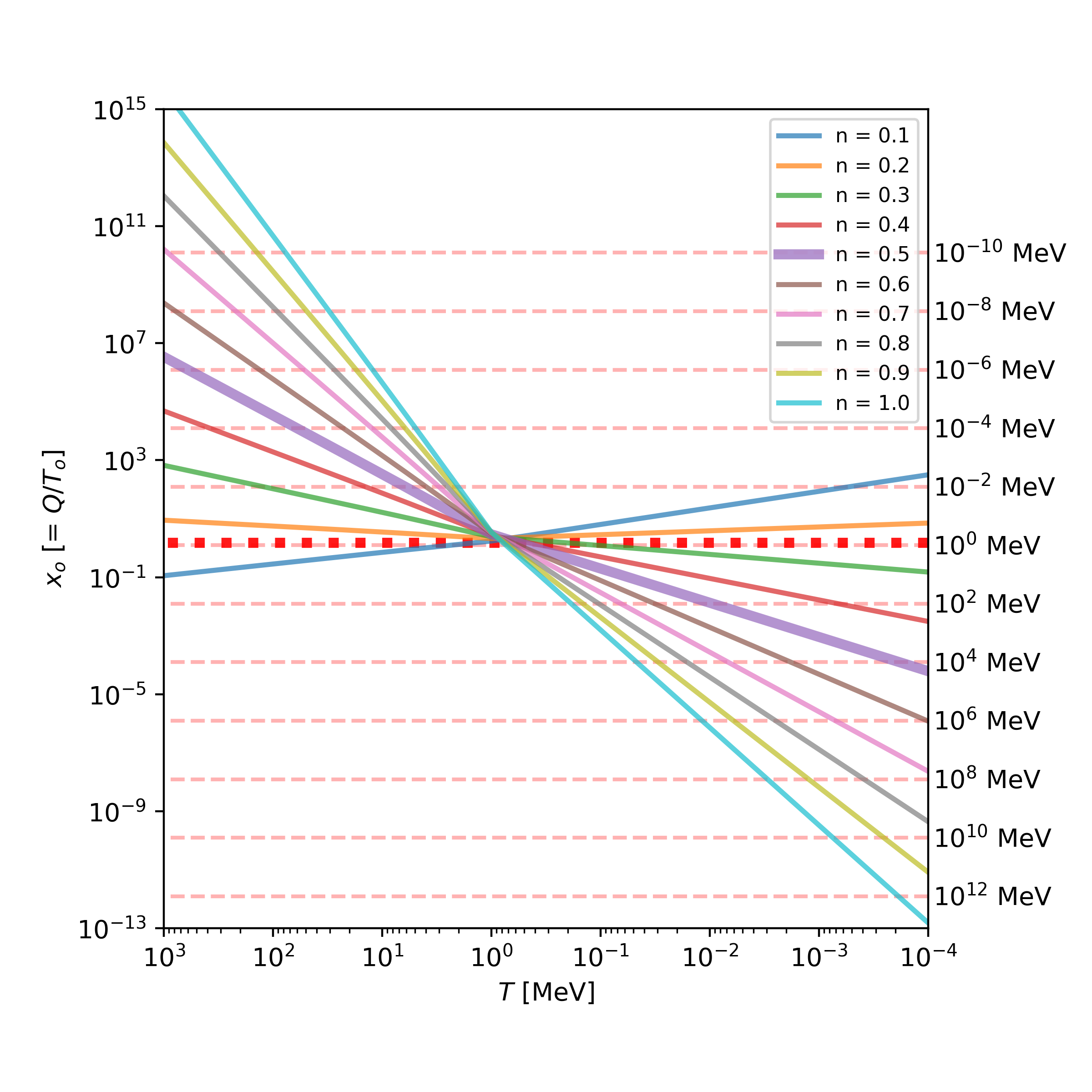}
\caption{\label{fig:i} Graphical representation of the inequality presented in Equation~\ref{ineq}. In this and all subsequent plots, temperature decreases to the right. The left-hand side depends only on the temperature at one second, represented as red horizontal lines in the above. The right-hand side of the inequality is shown as coloured lines, for a set of Universes with expansion index $n$. Nuclear statistical equilibrium for the proton to neutron ratio is maintained when a coloured line is \emph{above} a given pink dashed line. Our Universe is shown by the thick red dotted and thick purple solid lines.}
\end{figure}

In the following, we will parameterise the relationship between temperature and cosmic time as,
\begin{equation}
    T = T_o t^{-n}
\end{equation}
where $T_o$ is the temperature at a cosmic age of 1 second. This equation is accurate if the Universe's energy content is dominated by a single fluid. The standard cosmological model is radiation dominated through nucleosynthesis, which implies $n = 1/2$. Then, Equation~\ref{rate} can be written as,
\begin{equation}
    X_n = X_n^{eq} \left(
    1 + \frac{n}{\lambda_{np} \left( 1 + e^{-x} \right)^2 } \frac{x_o^{1/n}}{x^{1/n-1}}
    \right)
\end{equation}
where $x_o = \frac{Q}{T_o}$. Statistical equilibrium will be maintained so long as the second term in the parentheses is much smaller than one. We can illustrate this condition by separating the inequality as follows,
\begin{equation}
  x_o \ll \left[\left( \frac{255}{\tau_n} \right) \frac{x^{1/n - 6}}{n} (x^2 + 6x + 12) \left( 1 + e^{-x} \right)^2 \right]^n
  \label{ineq}
\end{equation}
The left-hand side depends on only the temperature of the Universe at 1 second, while the right-hand side is a function of temperature that depends on the expansion index $n$. In Figure~\ref{fig:i}, the horizontal red dashed lines represent the left-hand side, and the solid coloured lines represent the right-hand side of the inequality. Statistical equilibrium holds when a coloured line is above a given red dashed line.

For the standard cosmological model, dominated by radiation through nucleosynthesis,
\begin{equation}
T = \frac{ 0.8595 }{\sqrt{ t } } ~ \textrm{MeV}
\end{equation}
where $t$ is measured in seconds. This expansion is presented as a thick purple line in Figure~\ref{fig:i}. This line crosses the $T_o = 1~$MeV pink dashed line at around $T = 1~$MeV, so, as expected, neutrons and protons go out of equilibrium at a cosmic age of about $t\sim 1~$second.

To consider an alternative Universe, we have previously explored nucleosynthesis in the 
$R_h = ct$ Universe \citep{2016MNRAS.460..291L}. This model has linear expansion, and the relationship between the temperature and time is given by,
\begin{equation}
    T = \frac{1.036 \times 10^8}{ t } ~ \textrm{MeV}
\end{equation}
where the temperature is constrained by the Cosmic Microwave Background temperature today. In this Universe, that protons and neutrons go out of equilibrium at about $T \sim 5\times10^{-3}~$MeV, which implies a time scale of hundreds of years. As shown in \citet{2016MNRAS.460..291L}, at these comparatively low temperatures, ($\sim 6 \times 10^7~$K), the NSE abundances of heavier elements are non-negligible.

This section has illustrated the conditions for departure from NSE in an expanding Universe. The expansion of the Universe perturbs the abundances of the nuclear species away from their equilibrium values at a particular temperature and total density. Nuclear reactions drive the system back to equilibrium. Each of these processes can be characterised by a timescale, and the fastest process wins: either the rapid expansion of the Universe freezes the nuclear abundances in place (except for radioactive decay), or nuclear reactions maintain NSE.

\begin{figure}[tbp]
\centering 
\includegraphics[width=.5\textwidth]{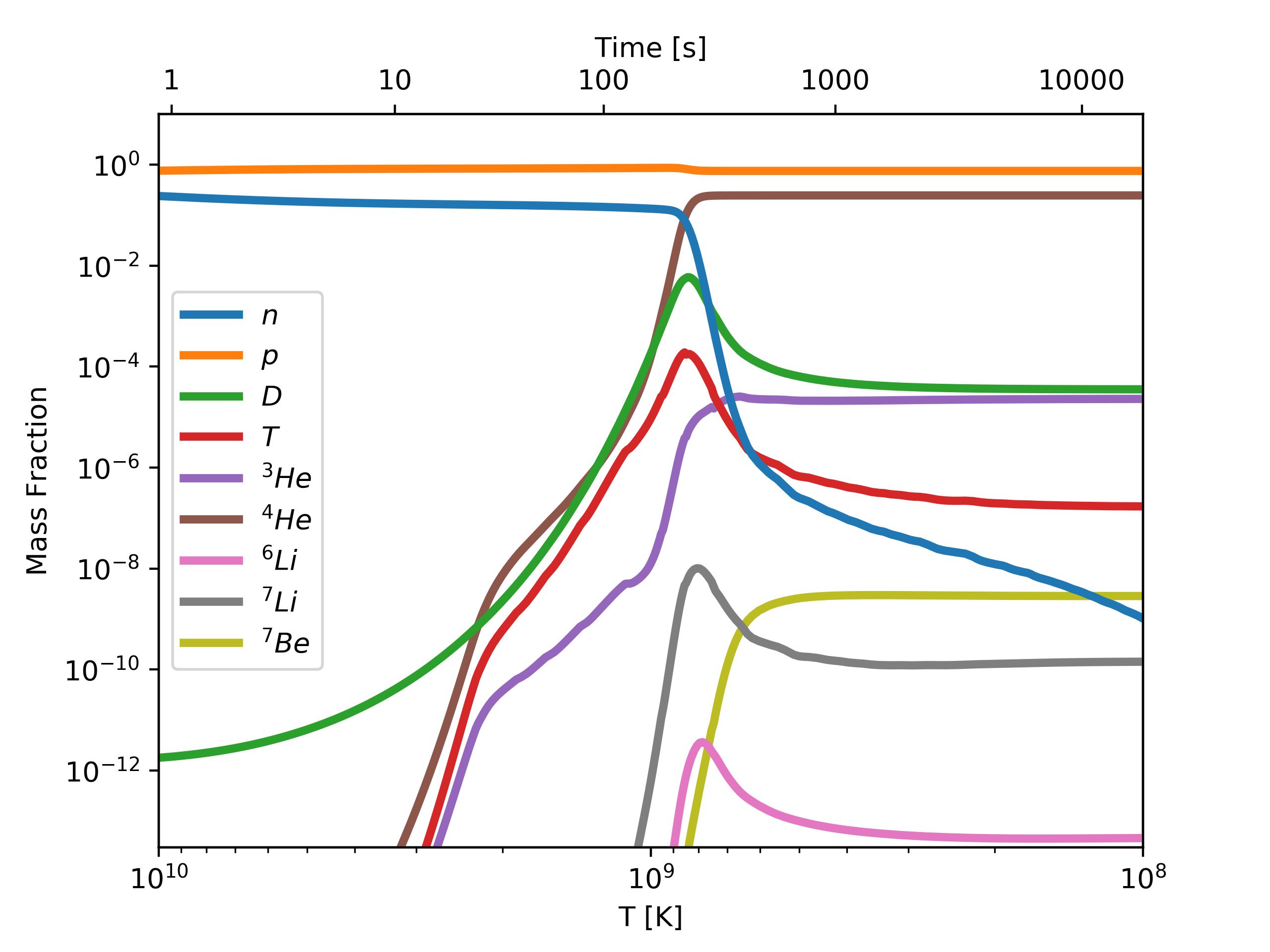}
\caption{\label{fig:BBN} Abundances of the light elements in our Universe as a function of temperature (bottom axis, decreases to the right) and time (top). As we will see in later sections, under conditions of NSE the abundance of deuterium (for example) peaks at $10^{-11}$; in our Universe, deuterium reaches abundances that are $10^8$ times higher than this. These nucelosynthesis pathways were integrated using {\it AlterBBN} \citep{2018arXiv180611095A}.}
\end{figure}

In our Universe, once freeze-out occurs at $T \approx 0.84~$MeV $= 9.7 \times 10^9~$K \citep{2004IJTP...43..669M}, the abundances of nuclear species can differ significantly from NSE. Big Bang Nucleosynthesis is far from a quasi-static, near-equilibrium process in our Universe. Figure \ref{fig:BBN} shows the abundances of the light elements in our Universe as a function of temperature and time, calculated by {\it AlterBBN} \citep{2018arXiv180611095A}. {\it AlterBBN} numerically integrates production and destruction rates over time for a complete nuclear reaction network that extends from protons and neutrons to \Oxs. As we will see in later sections, under conditions of NSE the abundance of deuterium (for example) peaks at $10^{-11}$; in our Universe, deuterium reaches abundances that are $10^8$ times higher than this, thanks to out-of-equilibrium nuclear reactions.

\section{Nuclear Statistical Equilibrium and Expansion}
\label{sec:nse}
\subsection{Nuclear Abundances in NSE}

This section will consider the effect of the expansion of the universe on Nuclear Statistical Equilibrium. We begin by briefly reviewing the principles of NSE, which have been expounded at length elsewhere \citep{1968psen.book.....C,KolbTurner,MukhanovCosm,IliadisNuclear}.

For non-relativistic matter in \emph{thermal} equilibrium, integrating the Maxwell-Boltzmann distribution over momentum gives the number density ($n_i$) of a particle species $i$ with particle mass $m_i$, degeneracy $g_i$ (also called the internal nuclear partition function) and chemical potential\footnote{As explained in \citet{2017ApJS..233...18L}, the chemical potential can be redefined as $\hat{\mu} = \mu - m_i - B_i$. This is particularly useful for numerical calculations of the NSE abundances.} $\mu_i$ at a temperature $T$,
\begin{equation} \label{eq:NSE}
n_i = g_i \left( \frac{m_i k T}{2 \pi \hbar^2}\right)^{\frac{3}{2}} \exp \left( \frac{\mu_i - m_i c^2}{k T}\right)
\end{equation}
The nuclear species $i$ has $Z_i$ protons and $N_i$ neutrons, giving atomic mass number $A_i = Z_i + N_i$. Hereafter, we will set the usual physical constants to unity: $c = \hbar = k = 1$.

If the reaction chain from neutrons and protons to species $i$ is in \emph{secular} equilibrium --- that is, if the rates of \emph{all} forward and reverse reactions are equal --- then the chemical potentials are related as:
\begin{equation}
\mu_i = Z_i \mu_p + N_i \mu_n ~.
\end{equation}
We define $\theta = (m_u T / 2 \pi)^{3/2}$, where $m_u = 1.6605$ g is the atomic mass constant, and the binding energy of species $i$ is $B_i = Z_i m_p + N_i m_n - m_i$. It is useful to express these number densities in terms of mass abundances relative to the total number density $n_b$ of protons and neutrons (inside and outside of nuclei). Then, given that $g = 2$ for the proton and neutron, we can combine the previous equations to give the mass abundances $X_i$,
\begin{align}
X_i &\equiv \frac{A_i ~ n_i}{n_b} \\
&= \frac{g_i ~ A_i^{\frac{3}{2}}}{2^{A_i}} \left( \frac{n_b}{\theta}\right)^{A_i-1} X_p^{Z_i} X_n^{N_i} \exp \left( \frac{B_i}{T}\right) \label{eq:XiNSE} \\ 
n_b &= \sum\limits_i A_i n_i
\end{align}
The relative abundances sum to one, and charge conservation ensures that that the number density of protons is equal to the net number density of electrons (that is, electrons minus positrons),
\begin{equation}\label{eq:YiNSE}
\sum\limits_i X_i = 1 \qquad \sum\limits_i \frac{Z_i}{A_i}X_i = Y_e ~.
\end{equation}
These two conditions close the set of equations, given $T$, $Y_e$ and $n_b$ (or equivalently, the mass density $\rho_b = m_u n_b$). The parameter $Y_e$ is related to the the ratio of total neutrons (inside and outside nuclei) to protons: $f_N = (1 - Y_e)/Y_e$. We integrate these equations using a modified version of the code of \citet{2008ApJ...685L.129S}\footnote{\url{http://cococubed.asu.edu/code_pages/nse.shtml}}.

In the context of big bang cosmology, the temperature and baryon number density $n_b$ are linked by the baryon-to-photon ratio,
\begin{align} 
\eta &\equiv \frac{n_b}{n_\gamma} \label{eq:ngam1} \\
n_\gamma &= \frac{2 \zeta(3)}{\pi^2} ~ T^3 \\
&= 3.37 \ten{4} \ro{ mol/cm}^3 ~ \left( \frac{T}{10^9 \ro{K}} \right)^3 ~, \label{eq:ngam2}
\end{align}
where $\zeta(3) \approx 1.202$ is the Riemann zeta function evaluated at 3.


In general, NSE will favour protons and neutrons at high temperatures, while at low temperatures, strongly-bound nuclei will be favoured, subject to the condition of fixed $Y_e$. It is not immediately obvious from Equation \eqref{eq:YiNSE} which nuclei will be most abundant, but is crucial to what follows and so we will briefly elaborate.

Suppose that at a given temperature, two species $i$ and $j$ dominate the relative abundances,
\begin{align*}
1 &\approx X_i + X_j \quad &&Y_e \approx \frac{Z_i}{A_i}X_i + \frac{Z_j}{A_j} X_j \\
\Rightarrow X_i &= \frac{Y_e - Z_j/A_j}{Z_i/A_i - Z_j / A_j} \quad &&X_j = \frac{Z_i/A_i - Y_e}{Z_i/A_i - Z_j / A_j}
\end{align*}
Note that, if species $i$ has the same proton/nucleon ratio as that required by $Y_e$, then $X_i = 1$; in other words, species $i$ can incorporate all the available protons and neutrons without any leftovers. More generally, if $Z_i / A_i < Y_e$, then $Z_j / A_k > Y_e$ (or vice versa). That is, if one species has a higher proton/nucleon ratio than required by $Y_e$, then the other must have a lower ratio.

\begin{figure*}[tbp]
\centering 
\includegraphics[width=\textwidth]{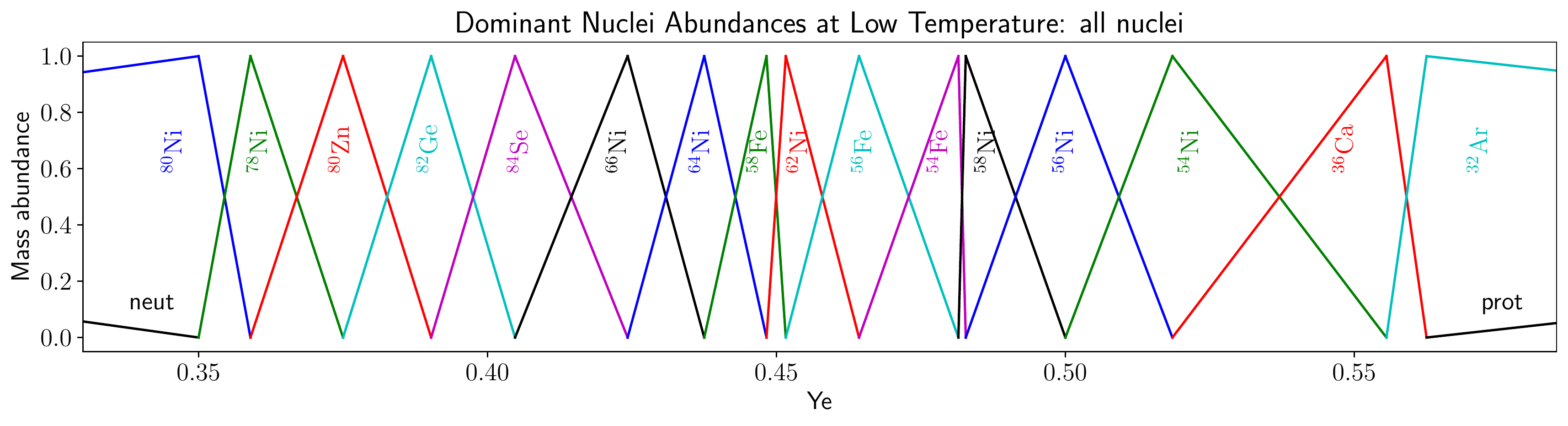}
\includegraphics[width=\textwidth]{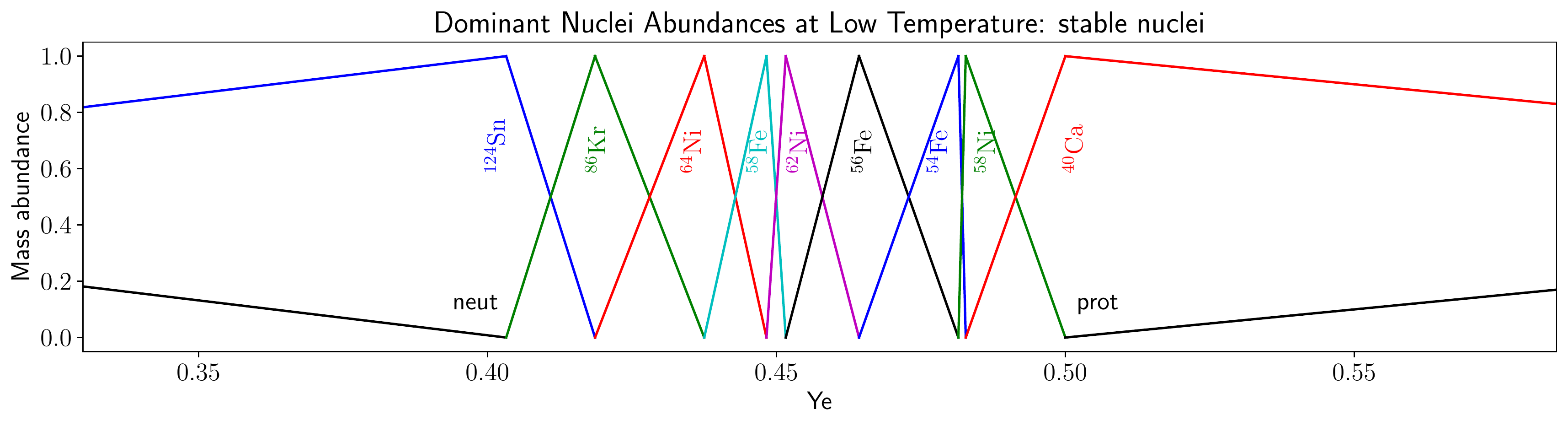}
\caption{\emph{Top}: The low-temperature NSE mass abundances of \emph{all} nuclei, as a function of the overall proton/nucleon ratio $Y_e$. This plot uses every known nuclear species in the NSDD database, including unstable ones. Each species ($^{80}$Ni, $^{78}$Ni, $^{80}$Zn, etc.) has a coloured line that matches its coloured text label. Each species dominates the abundance at the value of $Y_e$ that matches its own proton/nucleon ratio: $Y_e = Z_i/A_i$. In between these values, the abundance is shared between neighbouring species. \emph{Bottom}: As above, but using only nuclear species that are stable to all forms of radioactive decay.
\label{fig:NSE_Low_T} }
\end{figure*}

Given $X_i$ and $X_j$, we can solve for the mass abundances of the proton $X_p$ and neutron $X_n$ in Equation \eqref{eq:XiNSE}, and thus derive the abundance of any third species $k$. To be consistent with our assumption that two species dominate, it must be the case that $X_k \approx 0$.
\begin{itemize}
\item At high temperatures, the exponential term in Equation \eqref{eq:XiNSE} will tend to unity, and the dependence on temperature comes from the $\theta$ term: $X_k \propto T^{-\frac{3}{2}(A_k - 1)}$. Thus, all species with $A_k > 1$ will have vanishing abundance at large $T$, leaving only the proton and neutron.
\item At low temperature, the exponential term will dominate. If we create a 3D vector $\textbf{a}_i = (B_i,Z_i,N_i)$ for each nuclear species, then (after some algebra) the dependence of $X_k$ on the exponential term is,
\begin{equation} \label{eq:XkNSE}
X_k \propto \exp \left( \frac{1}{T} ~ \frac{\textbf{a}_k \cdot (\textbf{a}_i \times \textbf{a}_j)}{\left(Z_i N_j - N_i Z_j\right)} \right) ~,
\end{equation}
which uses the usual vector cross and dot products. To be consistent with our assumptions, the term in parentheses must be negative (and the denominator not equal to zero). Thus, for a given $Y_e$, we search for the pair of nuclei $i$ and $j$ such that a) $Z_i / A_i \leq Y_e$ and $Z_j / A_j \geq Y_e$, and b) for \emph{all} other species $k$, the term in parentheses above is negative.\footnote{Writing $f_i = N_i/Z_i$ and $b_i = B_i / Z_i$, we can characterise the available species as a cloud of points in $(f_i,b_i)$ space. We can write $Y_e$ equivalently as a neutron/proton ratio: $f_e = (1 - Y_e)/Y_e$. Then, equivalently and after even more algebra, we search for the pair of nuclei $i$ and $j$ such that a) $f_i \geq f_e \geq f_j$, and b) the line that passes through $(f_i,b_i)$ and $(f_j,b_j)$ passes above (larger values of $B/Z$) than the rest of the cloud. All things being equal, this will preference species with larger $B/Z$.} We have performed this search numerically for a set of nuclei from the Evaluated Nuclear Structure Data File (ENSDF)\footnote{Accessed via \url{https://www-nds.iaea.org/relnsd/NdsEnsdf/QueryForm.html}.}; the results are shown in Figure \ref{fig:NSE_Low_T}. The top panel shows all available nuclei, while the bottom panel shows only nuclei that are stable to all forms of radioactive decay. A set of highly-bound (large $B_i / A_i$) nuclei dominates the abundances at the value of $Y_e$ that matches their own proton/nucleon ratio: $Y_e = Z_i/A_i$. In between these values, the abundance is shared between neighbouring species. In particular, at small values of $Y_e$, free neutrons accompany the nucleus with the largest value of $B_i/Z_i$; at large values of $Y_e$, free protons accompany the nucleus with the largest value of $B_i/N_i$.
\item If we assume that the free neutron-proton ratio is maintained by the weak force in equilibrium, rather than the fixed ratio assumed above, then $X_n = X_p \exp(-Q_{np} / T)$, as in Equation \eqref{eq:nnnp}. In this case, some (simpler) algebra shows that at low temperature, the most abundant nucleus has the largest value of $B_i/A_i - Q_{np} N_i/A_i$, which is \Fe. (It is not, perhaps surprisingly, the most bound species per nucleon with the largest value of $B_i/A_i$, which is $^{62}$Ni.)
\end{itemize}

A potentially confusing aspect of NSE is that it takes no account of the stability of a nuclear species, beyond its binding energy. As we will see in the next section (Figure \ref{fig:NSE_eta}), small but non-zero abundances of famously highly-unstable elements such as \LiF\ (half-life: $4 \times 10^{-22}$ s) and \BeE\ (half-life: $8 \times 10^{-17}$ s) are present in NSE. In our Universe, the instability of these nuclei is one reason for BBN essentially ending at \HeF, and in stars it means that the nuclear path from protons to heavier elements requires a three-body interaction: the triple-alpha process. While \BeN\ is stable to all forms of decay, \BeE\ is more abundant in NSE because its binding energy per nucleon is larger.\footnote{More precisely, its larger value of $B_i/A_i - N_i Q_{np}/A_i$.}

How do such short-lived elements remain in equilibrium? An unstable element will be in equilibrium if its rate of decay is equal to its rate of formation. For example, \LiF\ and \BeE\ require:
\begin{equation} \label{eq:LiBe}
\HeF + p \rightleftharpoons \LiF \qquad \HeF + \HeF \rightleftharpoons \BeE ~.
\end{equation}
Equating the rates of production and decay, we find the conditions for secular equilibrium:
\begin{equation} \label{eq:ChemEq}
\begin{split}
\sv_{5} ~n_{\HeF} n_p = \lambda_{\LiF} n_{\LiF} \\ 
\frac{1}{2} \sv_{8} ~ n^2_{\HeF} = \lambda_{\BeE} n_{\BeE}
\end{split}
\end{equation}
where $\lambda_{\LiF}$ $~(\lambda_{\BeE})$ is the decay rate (in s$^{-1}$) for \LiF \ (\BeE), $\sv_{5}$ and $\sv_{8}$ are the thermally-averaged reaction rates per particle pair for the reactions in Equation \ref{eq:LiBe}, and $n_i$ is the number density of species $i$.

Both NSE \eqref{eq:NSE} and secular equilibrium \eqref{eq:ChemEq} can be rearranged to constrain, for example, the combination $n_{\HeF} n_p / n_{\LiF}$, given that equilibrium sets the chemical potentials: $\mu_{\LiF} = \mu_{\HeF} + \mu_p$. This might seem to overdetermine the abundances; how do the decays and reactions conspire to maintain both NSE and secular equilibrium? The answer is microreversibility \citep[][pg. 868]{messiah1968quantum}, otherwise known as the \emph{reciprocity theorem} \citep[][pg. 75]{IliadisNuclear}. Since nuclear interactions are time-reversal symmetric\footnote{The detection of $T$ violation in particle physics --- specifically, kaon interactions mediated by the weak force --- has lead to searches for detectable $T$-violations in low-energy nuclear physics. These tests, including experiments specifically aimed at reciprocity or detailed balance \citep{1983PhRvL..51..355B}, support the conclusion that nuclear reactions are invariant under time-reversal \citep[][pg. 75]{IliadisNuclear}. Any effect of $T$ violation from the weak force is negligible.}, there is a simple relationship between the forward and reverse decay / reaction rates per particle pair, which depends on the phase space available in the entrance and exit channels. The interaction doesn't care which state we call ``initial'' and which we call ``final''. For example, in the case of \BeE, the decay rate is related to the \HeF + \HeF\ reaction rate per particle pair as ($g_{\BeE} = g_{\HeF} = 1$),
\begin{equation} \label{eq:Bereact}
\frac{\frac{1}{2} \sv_{8}}{\lambda_{\BeE}} = \left( \frac{m_u T}{\pi} \right)^{\!\!-\frac{3}{2}} \exp \left( \frac{-m_{\BeE} + 2m_{\HeF}}{T}\right) ~.
\end{equation}
While the NSE relationship between the abundances holds only in equilibrium, this relationship between the decay / reaction rates per particle pair holds at all times. In equilibrium, both the right and left hand sides are equal to $n_{\BeE}/n^2_{\HeF}$. Perhaps counter intuitively, this implies that --- \emph{ceteris paribus} --- the rate of formation for an unstable nucleus increases with its decay rate.

\subsection{Reaction Rates Perturbed by Expansion}

The decay and reaction rates determine how quickly NSE is established from a given set of initial abundances. Given a reaction network that connects protons and neutrons to whichever nuclear species are most abundant, we can calculate the rate at which equilibrium will be established if the system is perturbed.

In particular, the expansion of the Universe (with scale factor $a$) will perturb both the number density of all species $n_i \propto a^{-3}$ and the reaction rates via the temperature $T \propto a^{-1}$. This section models a Universe that expands quasi-statically, that is, in very small increments with a sufficiently long pause at each step for the Universe to reestablish equilibrium. We ignore the relationship between expansion rate and energy density, given by the Friedmann equations, instead treating the expansion rate as an independent variable. This quasi-static scenario is very different to BBN in our universe.

Suppose that when the Universe has scale factor $a$, temperature $T$ and baryon number density $n_b$, a reaction $A + B \rightleftharpoons C + D$ of 4 non-relativistic species is in secular and thermal equilibrium. By microreversibility and Equation \eqref{eq:NSE}, the forward and reverse reaction rates are related at all times by:
\begin{align*}
\frac{\sv_{CD}}{\sv_{AB}} = f_r(T) \equiv \frac{g_A g_B}{g_C g_D} \left( \frac{m_A m_B}{m_C m_D} \right)^{\!\frac{3}{2}} \exp \left( -\frac{Q}{T} \right)
\end{align*}
where $Q = m_A + m_B - m_C - m_D$, and $f_r(T)$ depends on the reaction $r$, but is a reminder that temperature is the only dynamic variable on the right hand side. In secular equilibrium, the rate of change of the number density of species $A$ is zero,
\begin{align}
\dot{n}_C &= \sv_{AB} ~n_A n_B - \sv_{CD} ~n_C n_D \\
 &= \sv_{AB} \left[ n_A n_B - f_r(T) n_C n_D  \right] \\
 &= 0
\end{align}
Now suppose that the Universe instantaneously expands by a small amount $\dd a$, altering the temperature and number density. This will induce a small reaction rate $\dd \dot{n}_C$. The perturbed reaction rate can be shown to obey,
\begin{equation}
\frac{\dd \dot{n}_C} {\dd \log a} = \sv_{AB} ~ n_A n_B ~  \frac{Q}{T} ~.
\end{equation}
If $Q$ is positive, energy must be supplied for the reverse reaction to occur. As the Universe expands ($\dd a > 0$) and cools, the reaction products are energetically favoured. Thus, the net production rate of the reactant $C$ is positive.

If some of the reactants or products are relativistic, then instead of Equation~\eqref{eq:NSE}, we use Equation \eqref{eq:ngam2}, or its equivalent for fermions. At the relevant temperatures ($> 10^{9}$ K), neutrinos and electrons/positrons are relativistic \citep[][pg. 91]{MukhanovCosm}. For simplicity, we will assume that they are effectively massless and have zero chemical potential; see the extended discussion of this point in \citet[][pg. 89ff]{MukhanovCosm}. Following the same approach, the perturbed reaction rate is shown in Table \ref{tab:perturbreact} for a range of reaction forms.

\begin{table*}[ht]
\centering
\renewcommand{\arraystretch}{1.4}
\begin{tabular}{ |c|c|c| } 
 \hline
\multirow{2}{*}{Reaction} & Relativistic & \multirow{2}{*}{$\dd \dot{n}_C / \dd \log a$} \\
        & species & \\
 \hline
$A + B \rightleftharpoons C + D$ & none (or, $B$ and $D$) & $\sv_{AB} ~ n_A n_B ~ \frac{Q}{T}$ \\
$A + B \rightleftharpoons C + D$ & $D$ (e.g. photodisintegration) & $\sv_{AB} ~ n_A n_B \left( \frac{Q}{T} - \frac{3}{2} \right)$ \\
$A + B \rightleftharpoons C$ & none & $\sv_{AB} ~ n_A n_B \left( \frac{Q}{T} - \frac{3}{2} \right)$ \\ 
$A \rightleftharpoons B + C$ & none (e.g. $\alpha$ decay) & $\lambda_A ~ n_A \left( \frac{Q}{T} + \frac{3}{2} \right)$ \\ 
$A \rightleftharpoons B + C + D$ & $C$ and $D$ (e.g. $\beta$ decay) & $\lambda_A ~ n_A ~ \frac{Q}{T}$ \\
$A + B + C \rightleftharpoons D$ & none (e.g. triple $\alpha$) & $\sv_{ABC} ~ n_A~n_B~n_C ~ \left( \frac{Q}{T} - 3 \right)$ \\
 \hline
\end{tabular} 
\caption{Perturbed reaction rate for a variety of reaction forms, and with the relativistic species shown in the middle column. The quantity $Q$ is the total mass of reactants minus the total mass of products.}\label{tab:perturbreact}
\end{table*}

After the small expansion, the number density of species $i$ changes due to expansion at a rate $\dd n_i / \dd a$. We can also solve the NSE equations at the new temperature and density of the Universe, to find the new NSE abundances that have changed by $\dd n^{\ro{eq}}_{i} / \dd a$. Then,
\begin{equation} \label{eq:teq}
t_{eq,i} = \frac{\dd n^{\ro{eq}}_{i} / \dd a - \dd n_i / \dd a}{\dd \dot{n}_i / \dd a} ~,
\end{equation}
gives the timescale for species $i$ to return to equilibrium. Given that $T \propto 1/a$, we replace the derivative with respect to $\log a$ with the derivative with respect to $-\log T$. We consider only reactions that produce the larger nuclei when they are under-abundant; thus $t_{eq,i}$ is always positive.

We require a set of reactions that connect the relevant particles --- protons, neutrons, electrons/positrons, neutrinos --- with the largest nucleus that dominates elemental abundances in NSE, \Fe. Ideally, we would integrate all possible reactions until the system returned to equilibrium. However, we can simplify the calculation by considering the fastest reaction pathways. Further, given that our quasi-static calculation is only ever infinitesimally far from equilibrium --- and we thus do not specify a particular cosmological expansion history or temperature-density-time relation --- we use Equation \eqref{eq:teq}, rather than an explicit time integrator like {\it AlterBBN}. We add the equilibration times \eqref{eq:teq} for the production of each element in a given pathway in series, and add times for alternative pathways in parallel.

Ideally, we would integrate all possible reactions until the system returned to equilibrium. However, we can simplify the calculation by considering the fastest reaction pathway, and adding the equilibration times \eqref{eq:teq} for the production of each element in the pathway.

First, protons and neutrons interact via the weak force,
\begin{equation}
    p + e^- \rightleftharpoons n + \nu \qquad p + \bar{\nu} \rightleftharpoons n + e^+
\end{equation}
The reaction rates are calculated from the integrals in \citet[][pg. 100-1]{MukhanovCosm}.

For the production of nuclei, we use the ``21 isotope network'' of Cococubed\footnote{\url{http://cococubed.asu.edu/code_pages/burn_helium.shtml}}, described as the ``default workhorse network of MESA'' \citep{2011ApJS..192....3P}, which is a widely-used stellar evolution code. The temperature of BBN is comparable to the temperature of the cores of the most massive stars. Further, the baryon density of the universe is lower, so that no additional high-density (e.g. three-body) interactions are expected. Thus, we expect the MESA reaction chain to provide the appropriate reactions for our purposes. Reaction rates were drawn from the online database NETGEN\footnote{\url{http://www.astro.ulb.ac.be/Netgen/form.html}} \citep{2005A&A...441.1195A,2013A&A...549A.106X,2011ASPC..445..187X}. \HeF\ is produced by the following chain:
\begin{equation}
    \isotope[1]H(n,\gamma)\isotope[2]H(p,\gamma)\isotope[3]{He}(\isotope[3]{He},2p)\isotope[4]{He} ~.
\end{equation}
\HeF\ is connected to \Fe\  by an alpha-chain to \isotope[52]Fe, followed by free-neutron capture,
\begin{equation}
\begin{split}
&\isotope[4]{He}(2\alpha,\gamma) 
\isotope[12]{C}(\alpha,\gamma) 
\isotope[16]{O}(\alpha,\gamma) 
\isotope[20]{Ne}(\alpha,\gamma)\isotope[24]{Mg} \ldots \\ 
&\isotope[24]{Mg}(\alpha,\gamma)
\isotope[28]{Si}(\alpha,\gamma)
\isotope[32]{S}(\alpha,\gamma)
\isotope[36]{Ar}(\alpha,\gamma)\isotope[40]{Ca} \ldots \\ 
&\isotope[40]{Ca}(\alpha,\gamma)
\isotope[44]{Ti}(\alpha,\gamma)
\isotope[48]{Cr}(\alpha,\gamma)
\isotope[52]{Fe}(n,\gamma)\isotope[53]{Fe} \ldots \\ 
&\isotope[53]{Fe}(n,\gamma)
\isotope[54]{Fe}(n,\gamma)
\isotope[55]{Fe}(n,\gamma)
\isotope[56]{Fe} ~.
\end{split}
\end{equation}
In addition, we include the effect of the following parallel pathways.

\begin{itemize}
    \item \HeF\ can also be produced via neutron capture  $\isotope[3]{He}(n,\gamma)\isotope[4]{He}$, and via tritium $\isotope[2]H(n,\gamma)\isotope[3]H(p,\gamma)\isotope[4]{He}$; tritium has a lifetime of 12.32 years, long enough to be measured in underground water sources \citep{JB1971,JB1973}. 
    \item For each reaction in the ``alpha-chain'' between \isotope[24]Mg and \isotope[52]Fe, there is an alternative two-step reaction that involves a proton catalyst, e.g. $\isotope[24]{Mg}  (\alpha,p)\isotope[27]{Al}(p,\gamma)\isotope[28]{Si}$.
    \item The reactions $\isotope[12]C + \isotope[12]C \rightleftharpoons \isotope[20]{Ne} + \alpha$, $\isotope[12]C + \isotope[16]O \rightleftharpoons \isotope[24]{Mg} + \alpha$ and $\isotope[16]O + \isotope[16]O \rightleftharpoons \isotope[28]{Si} + \alpha$ are included.
    \item We include an alternative pathway in the Cococubed network between \isotope[54]Fe and \Fe\  via \isotope[57]Co.
    \item The alpha-chain becomes slower as we approach \isotope[52]Fe, as it can create only nuclei with $Z = N$ and these elements become less bound: \isotope[44]Ti, \isotope[48]Cr, and \isotope[52]Fe are radioactive. The slowest of the alpha-chain reactions over much of the relevant temperature range is $\isotope[44]{Ti}(\alpha,\gamma)\isotope[48]{Cr}$. We consider an alternative pathway that involves beta-decay and electron capture: $\isotope[44]{Ti}(e^-,\nu)$$\isotope[44]{Sc}(,e^+ \nu)$$\isotope[44]{Ca}(\alpha,\gamma)$$\isotope[48]{Ti}(\alpha,\gamma)$ $\isotope[52]{Cr}(\alpha,\gamma)$$\isotope[56]{Fe}$.
\end{itemize}

\subsection{Maintaining NSE in an Expanding Universe}

\begin{figure*}[tbp]
\centering 
\includegraphics[width=.495\textwidth]{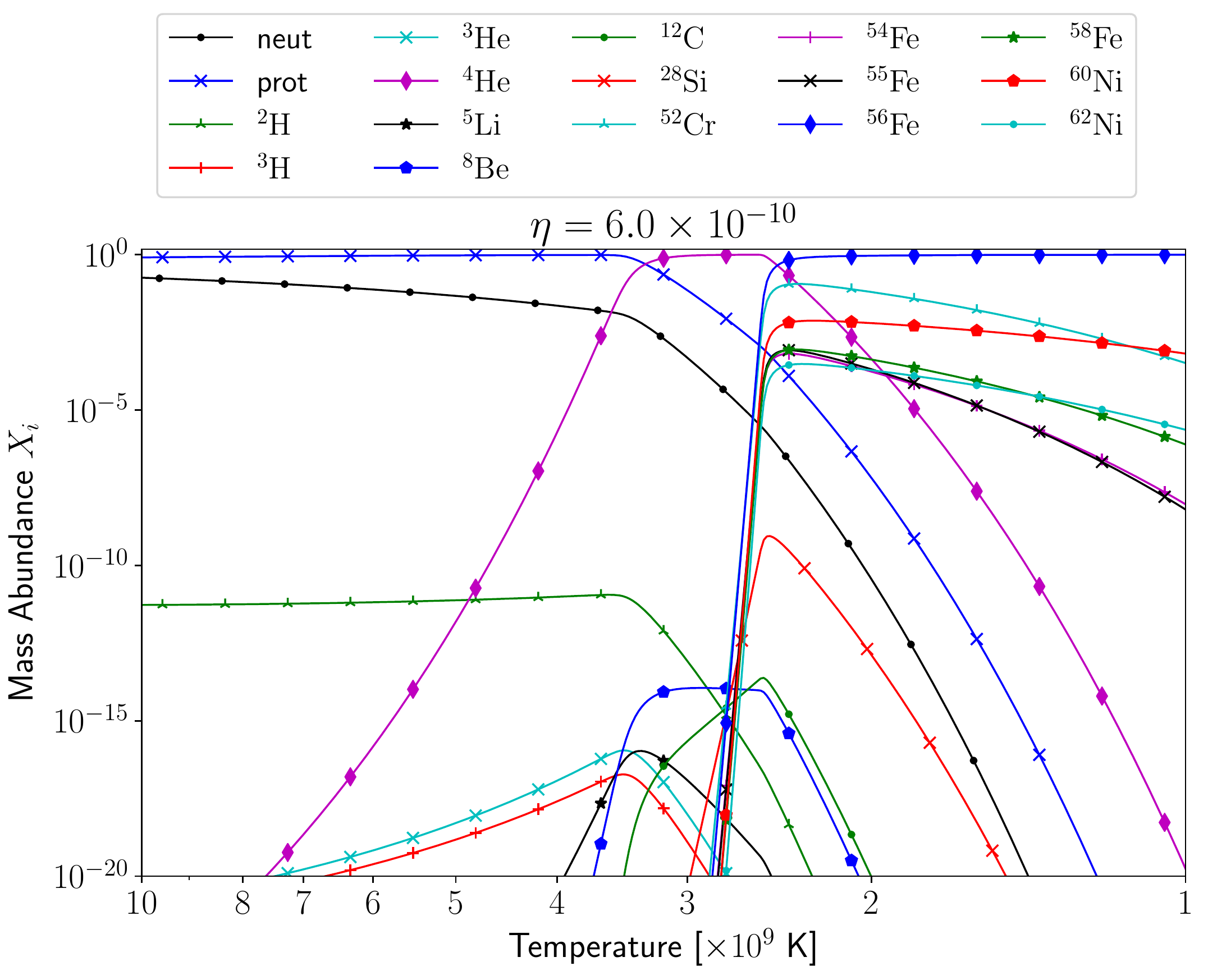}
\includegraphics[width=.495\textwidth]{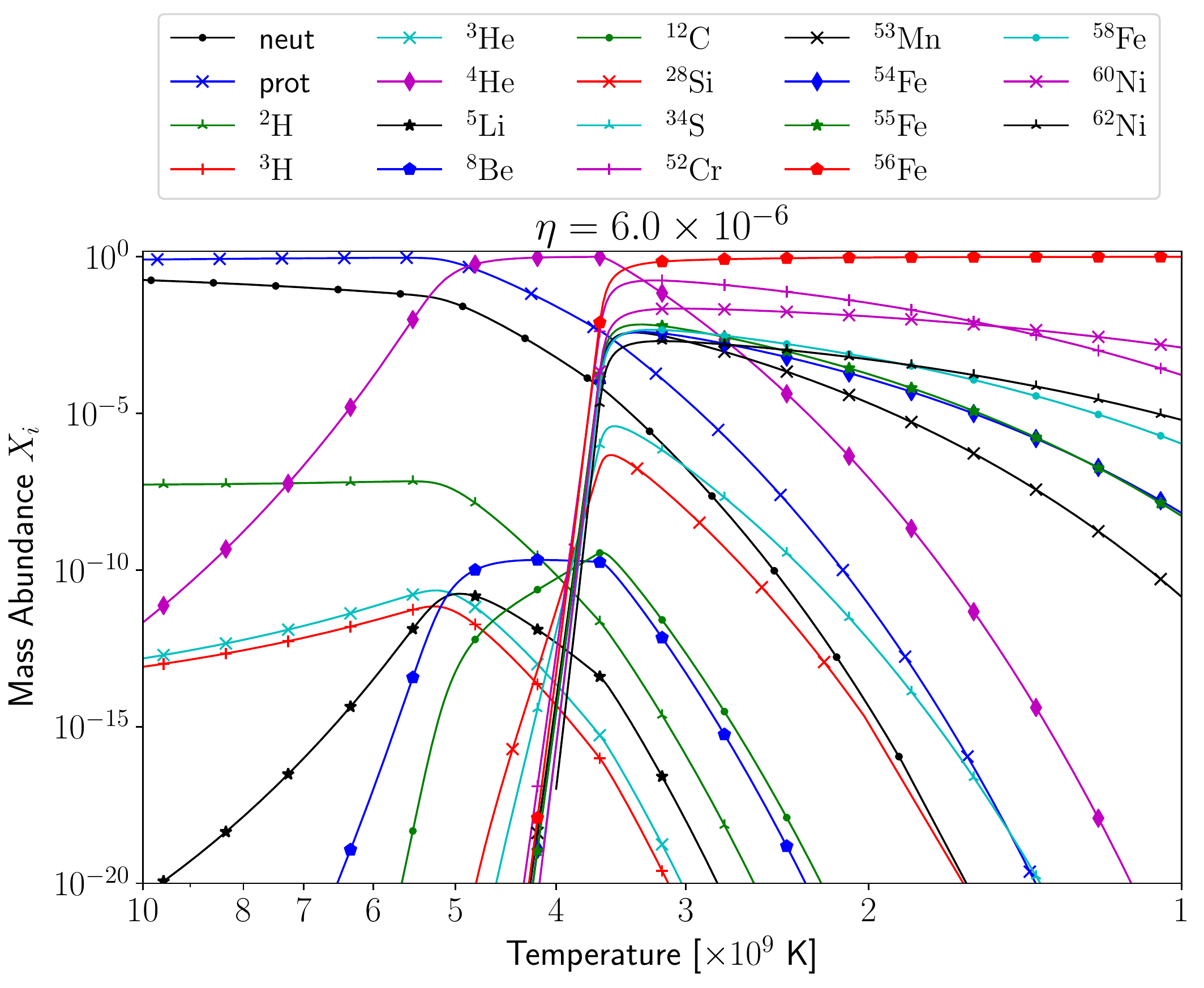} \\
\includegraphics[width=.495\textwidth]{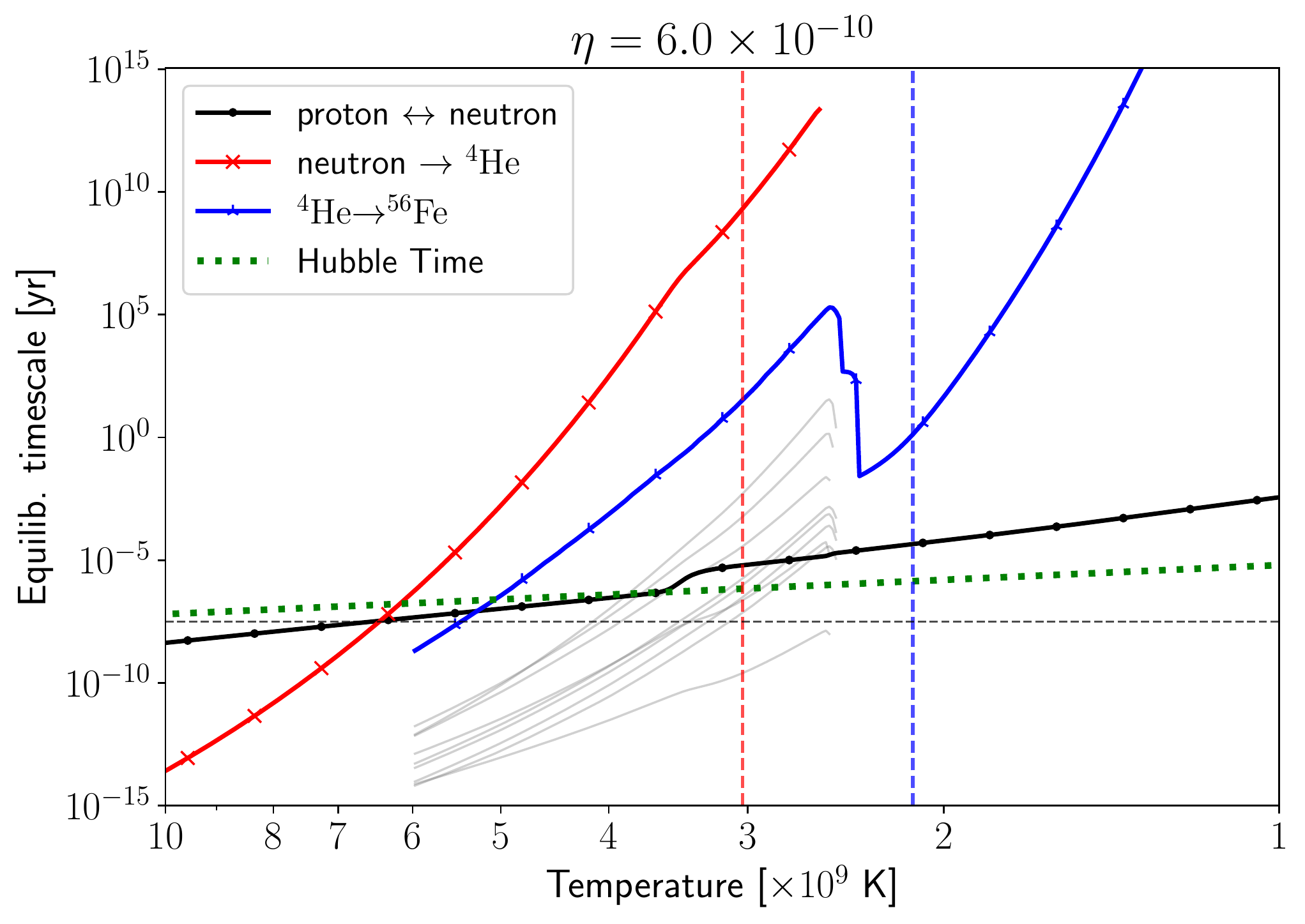}
\includegraphics[width=.495\textwidth]{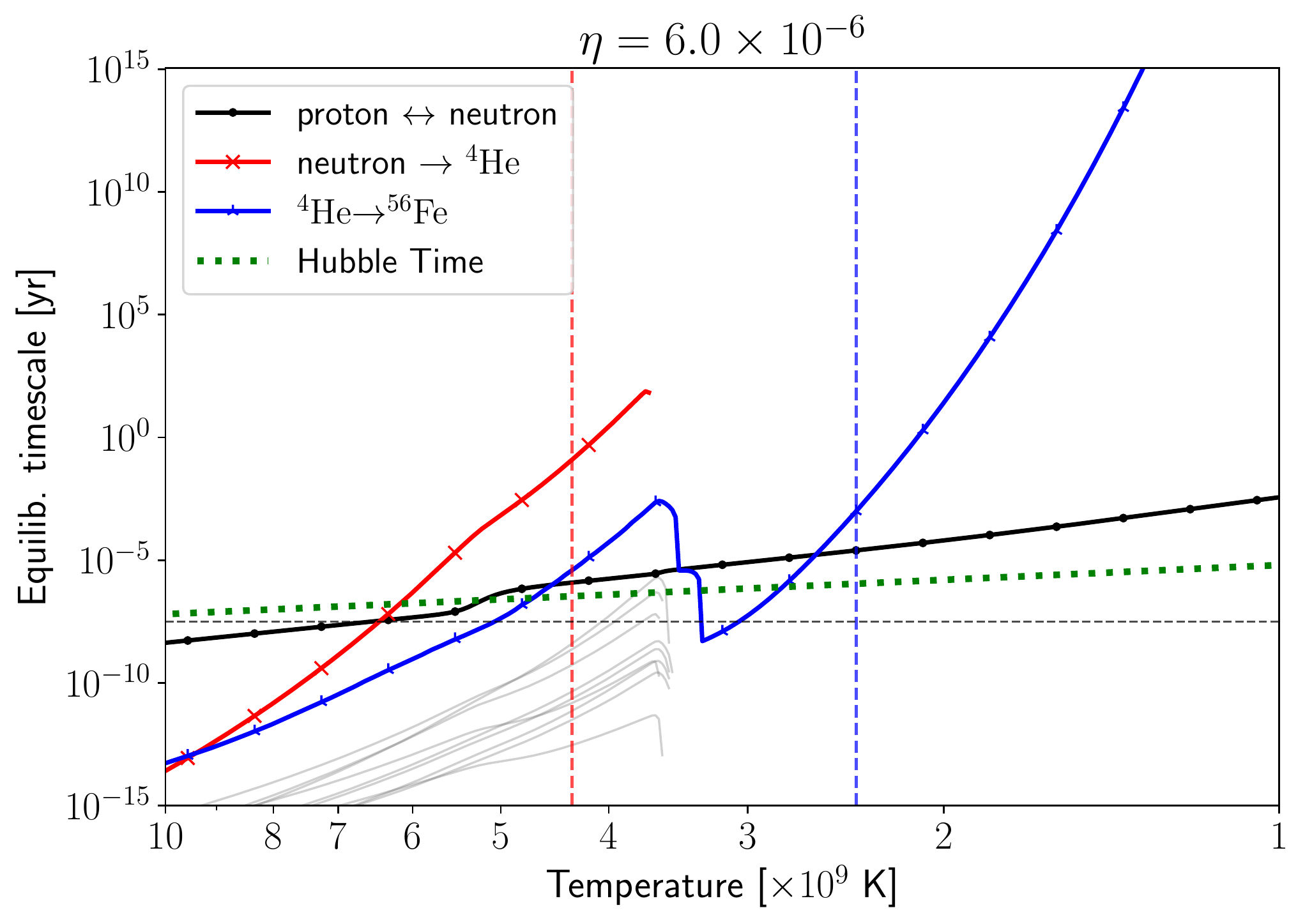}
\caption{
\emph{Top Panels:} NSE mass abundances as a function of temperature (decreasing to the right) for a Universe with baryon-to-photon ratio $\eta$. The top left panel is for the value of $\eta = 6 \times 10^{-10}$, as in our Universe, and the right panel shows the effect of increasing the baryon-to-photon ratio to $\eta = 6 \times 10^{-6}$. Note that the line styles are not the same for the two panels. To reduce clutter, not all nuclei are shown: the plotted nuclei are those that, at some temperature, are in the top five abundances. Both panels show a similar pattern: protons and neutrons at high temperature, \HeF\ coming to dominate at intermediate temperatures, and highly bound nuclei (particularly \Fe) dominating at low temperatures.
\emph{Bottom Panels:} The equilibration timescales from Equation \eqref{eq:teq} for the Universe with $\eta = 6 \times 10^{-10}$ (left) and $\eta = 6 \times 10^{-6}$ (right). As shown in the legend, the solid lines are for the equilibrium between protons and neutrons, from neutrons to \HeF, and from \HeF\ to \Fe. (Neutrons are always overabundant, and thus their production is not a rate-limiting step. The neutron equilibration time is shown for comparison with freeze-out in our Universe.) The horizontal dashed line shows $t_{eq} = 1$ second. The dotted green line shows $1/H$, the Hubble time, from Equation \eqref{eq:HubbleTime}. The dashed vertical lines show the temperature at which the \HeF\ (left, red) and \Fe\  (right, blue) have NSE abundances greater than 0.9. The thin grey lines show the many reactions in the chain from \HeF\ to \Fe\ --- because they span many orders of magnitude, the overall $t_{eq}$ is approximately equal to the timescale of the slowest reaction. In both panels, the production of \HeF\ is the rate-limiting step.
}
\label{fig:NSE_eta} 
\end{figure*}

Figure \ref{fig:NSE_eta} (top) shows the NSE mass abundances as a function of temperature (decreasing to the right) for a Universe with a fixed baryon-to-photon ratio ($\eta$), and with $X_n = X_p \exp(-Q / T)$. The left panel is for the value of $\eta = 6 \times 10^{-10}$, as in our Universe. Beginning at high temperatures, the Universe is dominated by protons and neutrons, with a small number of deuterons. As the temperature decreases, \HeF  comes to dominate the abundances, due to its anomalously-large binding energy per nucleon amongst small elements. Small amounts of other light elements are also present, including unstable species. At about $2.7 \times 10^{9}$ K, the abundances of highly-bound elements (chromium, iron, nickel) rises rapidly. As expected, at small temperatures, \Fe\ is the most abundant nucleus.

The top right panel of Figure \ref{fig:NSE_eta} shows the effect of increasing the baryon-to-photon ratio to $\eta = 6 \times 10^{-6}$. The qualitative features are very similar. The lower ratio of photons reduces the photo-disintegration rate, allowing nuclei species to exist at greater abundances at larger temperatures. The transition to larger, bound nuclei happens at higher temperatures, with \Fe\ again becoming the most abundant nucleus at low temperatures. If we decrease $\eta$, the pattern makes a similar shift to the lower temperatures.

The lower panels of Figure \ref{fig:NSE_eta} show the equilibration time (Equation \ref{eq:teq}) for a given value of $\eta$, for the reaction network outlined above. As shown in the legend, the solid lines are for the equilibrium between protons and neutrons, from neutrons to \HeF, and from \HeF\ to \Fe. The horizontal line shows $t_{eq} = 1$ second. The dashed vertical lines show the temperature at which the \HeF\ (left, red) and \Fe\ (right, blue) have NSE abundances greater than 0.9. The dotted green line shows the Hubble time, relevant to the next section,
\begin{align} 
t_H \equiv H^{-1} &= \left( \frac{8 \pi G}{3} \kappa \right)^{\!\!-\frac{1}{2}} T^{-2} \\
&= 198.9 \left( \frac{T}{10^9 \ro{ K}}\right)^{-2} \ro{ s} \label{eq:HubbleTime} \\
\ro{where} \quad \kappa &= \frac{\pi^2}{30} ~ \left(g_b + \frac{7}{8} g_f \right) ~,
\end{align}
where the degeneracy factor for bosons is $g_b = 2$ (2 photon polarisation states), and for fermions is $g_f = 10$ (3 neutrino species + 3 antineutrinos + 2 electron spin states + 2 positron spin states) at the relevant temperatures \citep[][pg. 94]{MukhanovCosm}. The thin grey lines show the many reactions in the chain --- because they span many orders of magnitude, the overall $t_{eq}$ is approximately equal to the timescale of the slowest reaction. The lines stop at low temperature because the NSE abundance begins to decline; the species is over-abundant, and so the equilibration timescale no longer depends on the rate of production of the species. \emph{Note well} that, in this quasi-static case, neutrons are always overabundant, and thus their production is not a rate-limiting step. The neutron equilibration time is shown for comparison with freeze-out in our Universe.

For $\eta = 6 \times 10^{-10}$ (as in our Universe, bottom left panel), the timescale to \HeF\ increases rapidly as the temperature drops and its reactants (tritium and \isotope[3]He) have very small abundances. To make a Universe that is at least 90\% helium by mass at some temperature, the Universe must expand on a timescale of $\sim 2 \times 10^{9}$ years at $\sim 3 \times 10^{9}~$K.

Once \HeF\ has been produced, the alpha-chain reactions can produce \Fe\ on a timescale of $2 \times 10^5$ years. This time is set by the ``bump'' in the equilibration timescale at $2.4 \times 10^{9}$ K, which is caused by a transition between the neutron pathway to \Fe\ and the alternative pathway via beta-decay and electron capture.

The bottom right panel of Figure \ref{fig:NSE_eta} shows the the equilibration time for the larger value of $\eta = 6 \times 10^{-6}$. The qualitative features are similar. Because the abundance of \HeF\ and \Fe\ dominates at larger temperatures, where the reaction rates are larger, the respective timescales are shorter. \HeF\ reaches 90\% abundance by mass after $\sim 0.12$ years (45 days), at $4.3 \times 10^9~$K. At lower temperatures, \Fe\ reaches 90\% abundance by mass after a further $\sim 23$ hours. Smaller values of $\eta$ show the opposite trend --- lower temperatures, slower reactions, longer equilibration timescales.

\begin{figure}[tbp]
\centering 
\includegraphics[width=.5\textwidth]{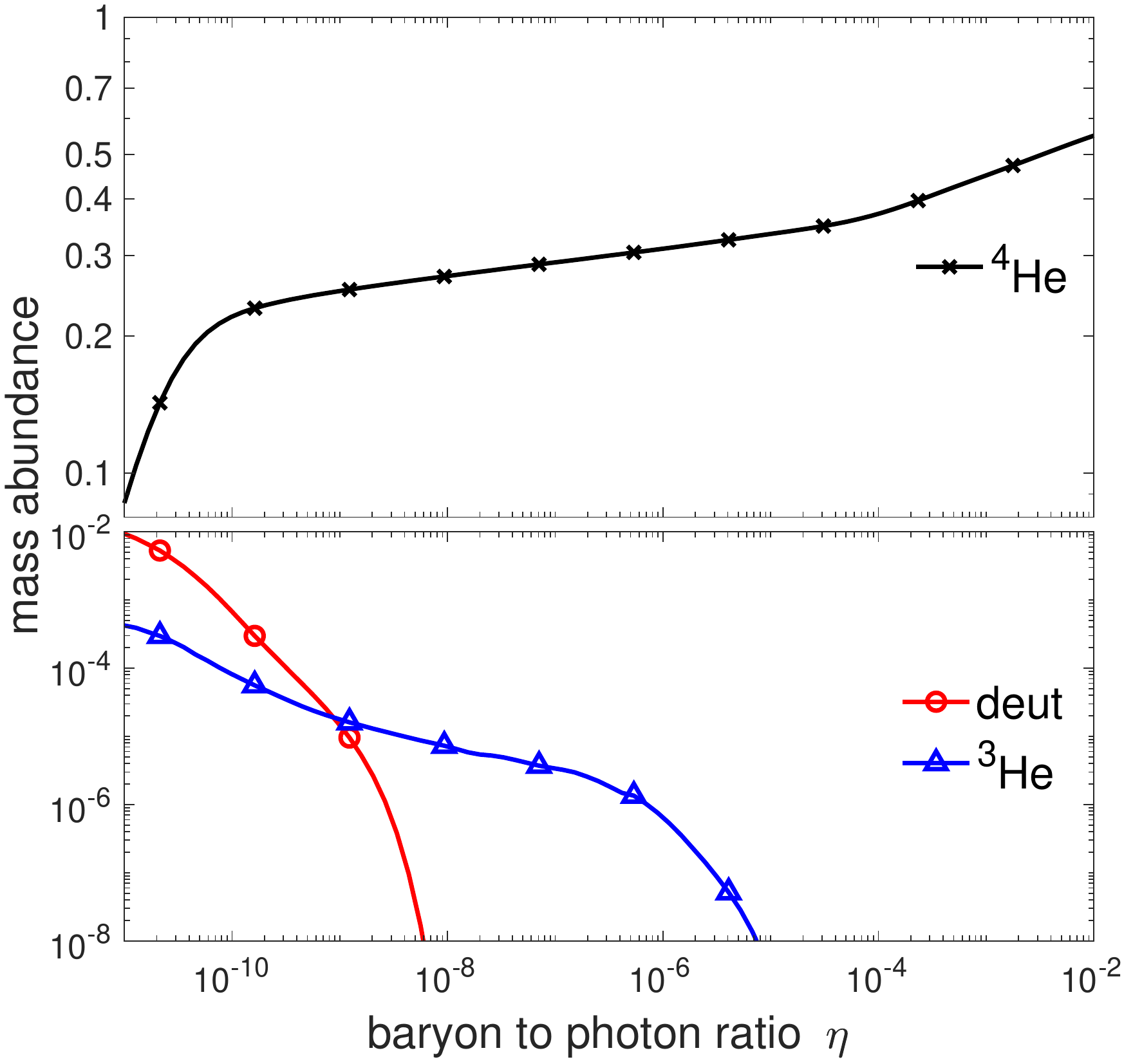}
\caption{Late-time elemental abundances for \HeF, deuterium and \isotope[3]He as a function of the baryon to photon ratio $\eta$, as calculated by the nucleosynthesis code {\it AlterBBN} \citep{2018arXiv180611095A}. As $\eta$ increases, there are fewer photo-disintegrating photons, so species can survive at higher temperatures. NSE abundances peak at higher temperatures, and the rates of NSE-establishing rates are faster. As a result, the abundance of \HeF increases with the baryon-to-photon ratio.
\label{fig:AlterBBN_eta} }
\end{figure}

We can illustrate the trend with $\eta$ using the {\it AlterBBN} nucleosynthesis code \citep{2018arXiv180611095A}. Though the reaction chain extends only to $\isotope[16]O$, the production of \isotope[4]He illustrates the trend above. Figure \ref{fig:AlterBBN_eta} shows the late-time elemental abundances for deuterium, \isotope[3]He and \HeF. As $\eta$ increases, there are fewer photo-disintegrating photons, so species can survive at higher temperatures. NSE abundances peak at higher temperatures, where the rates of NSE-establishing rates are faster. As a result, the abundance of $\isotope[4]He$ increases with $\eta$, as expected\footnote{AlterBBN seems to encounter numerical issues above $\eta = 10^{-2}$.} from Figure \ref{fig:NSE_eta}. 

Summarising, for a Universe with $\eta = 6 \times 10^{-10}$ (like ours) to have fused a substantial fraction of its baryons into \Fe\ , it would have had to cool from $\sim 10^{10}~$K to $\sim 2 \times 10^{9}~$K over a period of about a billion years. This is --- comfortably --- longer than the minute or so that this cooling period lasted in our Universe. The rate limiting step turns out to be the production of \HeF, due to the very small NSE abundances of its reactants, tritium and \isotope[3]He.

\section{How to Make an Iron Universe}
\label{sec:ironUniverse}
NSE can be maintained only when cosmological expansion is slow compared to the reaction rates that maintain equilibrium. Can we make a Universe that burns all the way to \Fe?

We have little scope for making an iron Universe by slowing the expansion of the Universe. Consider an expanding Universe that includes radiation (with density $\rho_r$), matter ($\rho_m$), a cosmological constant ($\rho_\Lambda$), a curvature term ($-kc^2/R^2$) and an extra form of energy ($\rho_X$) with equation of state $w = P_X/\rho_X$. For each of these components in the Friedmann equation for the expansion rate of the Universe, we can define characteristic times: $t_r = (8 \pi G \rho_r/3)^{-1/2}$ etc., and $t_c = R/c$. Then,
\begin{equation}\label{eq:Hubble_t}
    H^2 = \frac{1}{t_r^2} + \frac{1}{t_m^2} +\frac{1}{t_X^2} +\frac{\ro{sign}(\rho_\Lambda)}{t_\Lambda^2} - \frac{k}{t_c^2} ~.
\end{equation}
If we want the Universe to expand quasi-statically between two temperatures on a timescale $t_{eq}$ --- recalling that $1/H$ is a characteristic timescale for expansion --- we need $H < 1/t_{eq}$ between the two temperatures. However, at a temperature of $3 \times 10^9$ K, the energy density of radiation (relativistic species) implies that $t_r \approx$ a few minutes. Additional matter and ``X'' terms do not help: adding energy density can only make $H$ larger. We can cancel the positive contributions in Equation \eqref{eq:Hubble_t} with a negative cosmological constant, but this is a version of Einstein's static universe: it cannot be made to subsequently expand and cool. If it departs from $H = 0$, it can only recollapse.

The best we can do is a ``stalling'' Universe: if we include positive curvature and a positive cosmological constant, we can arrange for the Universe to stall, with $H \approx 0$ as the Universe almost recollapses under the attractive pull of matter and radiation, only to be saved by the repulsion of the cosmological constant. However, because each of the terms in Equation \eqref{eq:Hubble_t} varies differently with the scale factor $a$, the Universe must stall at a particular value of $a$, that is, a particular temperature. This is not quasi-static expansion, but would still allow reactions to proceed. This scenario is extremely fine-tuned --- the terms in Equation \eqref{eq:Hubble_t}, each of order (a few minutes)$^{-2}$, would have to fortuitously cancel to many decimal places to stall the Universe for the required nuclear timescales.

However, as shown in the previous section, we can effectively speed up the nuclear reactions by increasing the baryon-to-photon ratio $\eta$. This shifts the NSE abundances to higher temperatures, where the reaction rates are substantially larger.

\begin{figure*}[tbp]
\centering 
\includegraphics[width=.495\textwidth]{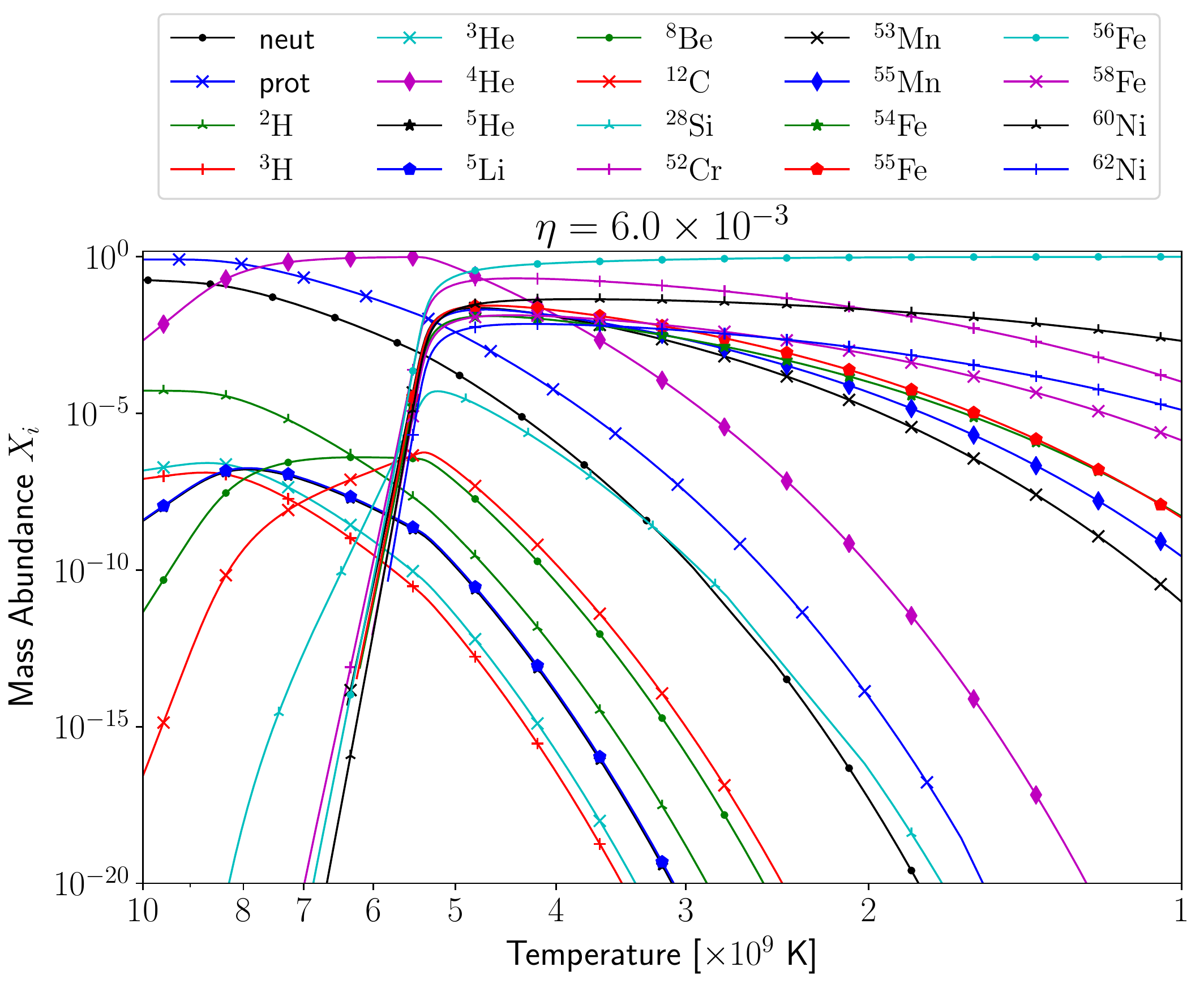}
\includegraphics[width=.495\textwidth]{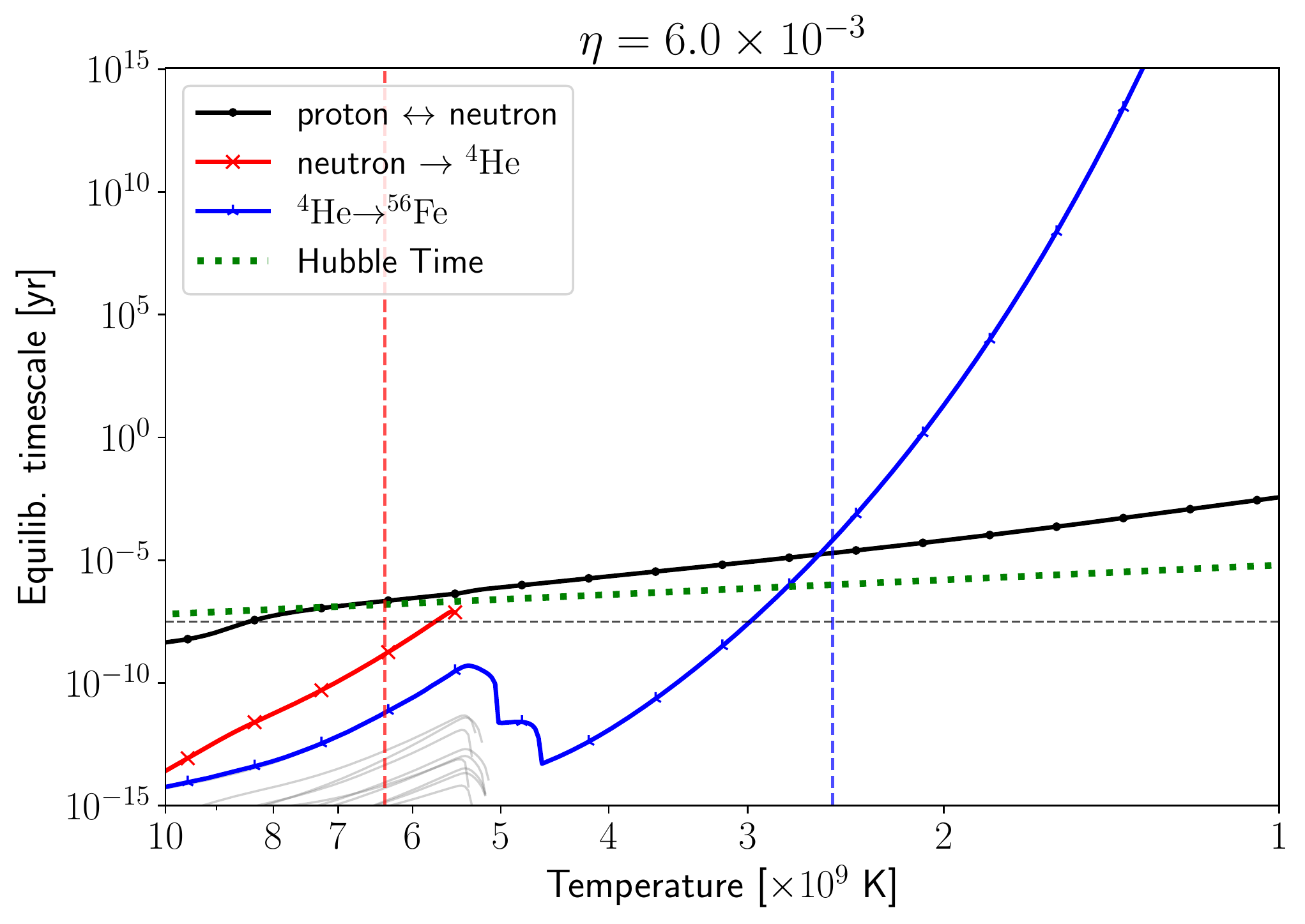}
\caption{The same as Figure \eqref{fig:NSE_eta}, but for a Universe with $\eta = 6 \times 10^{-3}$. The dotted green line in the right panel shows $1/H$, the Hubble time, from Equation \eqref{eq:HubbleTime}. The \HeF\ and \Fe\ equilibration times in the right panel remain below the Hubble time for temperatures above $2.8 \times 10^9~$K, meaning that these reactions remain in equilibrium. At $2.8 \times 10^9~$K, the NSE abundance of \Fe\ is $X = 0.87$.
}
\label{fig:NSE_eta_3} 
\end{figure*}

Figure \ref{fig:NSE_eta_3} shows the NSE abundances (left) and equilibration timescale (right) for a Universe with $\eta = 6 \times 10^{-3}$. The dotted green line in the right panel shows $1/H$, the Hubble time, from Equation \eqref{eq:HubbleTime}. The \HeF\ and \Fe\ equilibration times remain below the Hubble time for temperatures above $2.8 \times 10^9~$K, meaning that these reactions remain in equilibrium. At $2.8 \times 10^9~$K, the NSE abundance of \Fe\ is $X = 0.87$. A Universe with sufficiently large baryon-to-photon ratio will burn a significant fraction of its baryonic content into \Fe.

\section{Discussion}

In this section, we discuss the approximations made in our calculation. Our calculation considered a limited number of pathways between protons/neutrons and highly-bound nuclei such as iron. We assume that Cococubed and MESA have identified the fastest reaction path, or at least, there is no alternative path that is significantly faster.

We have not integrated the reaction network to calculate abundances, but instead have relied on timescales calculated from equilibrium reaction rates perturbed by the expansion of the universe. This approach is appropriate to the scenario we consider, in which the abundances stay close to their equilibrium value until the large elements are produced. Our conclusions are drawn in light of this order-of-magnitude calculation.

Our assumption of a Maxwell-Boltzmann distribution for non-relativistic species (Equation \ref{eq:NSE}) and $n \propto T^3$ for relativistic species breaks down if any of the relevant species becomes degenerate. This depends on the chemical potentials of the relevant species. As shown in detail in \citet[][pg. 91ff.]{MukhanovCosm}, in the Lepton era --- when electrons are relativistic but protons are not --- the ingredients of the universe are photons, leptons ($e$, $\mu$, $\tau$), neutrinos ($\nu_e$, $\nu_\mu$, $\nu_\tau$), light baryons ($p$, $n$, $\Lambda$), and mesons ($\pi^0$, $\pi^\pm$). Taking into account relations between the chemical potentials due to reactions such as $\mu^- \rightarrow e^- + \bar{\nu}_e + \nu_\mu$, only five are independent, which can be taken to be $\mu_{e^-}$, $\mu_{n}$, $\mu_{\nu_e}$, $\mu_{\nu_\mu}$, and $\mu_{\nu_\tau}$.

These five chemical potentials are in turn specified by five conserved quantities: the baryon-to-entropy ratio $B$, net electric charge-to-entropy ratio $Q$, and three lepton number-to-entropy ratios $L_e$, $L_\mu$, and $L_\tau$. The entropy decreases with the scale factor as $a^{-3}$, and is dominated by relativistic species: $s \sim n_\gamma \sim T^3$. It follows that $B \sim \eta$.

For non-relativistic species, the Maxwell-Boltzmann distribution is an accurate approximation at the relevant temperatures if $m \gg T$ and $(m - \mu) \gg T$. For the proton, the mass (938 MeV) is significantly larger than the temperature of BBN ($\sim$ MeV). Further, \citet[][pg. 93]{MukhanovCosm} shows that:
\begin{equation}
    \frac{m_p - \mu_p}{T} \approx \ln \left(\frac{1}{B} \left( \frac{m_p}{T} \right)^{3/2} \right) \approx \ln \left(\frac{1}{B} \right) + 10 ~.
\end{equation}
Hence, even for $B \sim \eta \sim 1$, the Maxwell-Boltzmann distribution is sufficiently accurate.

For relativistic species, the relation $n \propto T^3$ is accurate if $|\mu| \ll T$. For photons, $\mu_\gamma = 0$. The most important relativistic species during BBN is the electron. As shown in \citet[][pg. 93]{MukhanovCosm}, the electric neutrality of the universe implies that: 
\begin{equation}
    \frac{|\mu_e|}{T} \approx B .
\end{equation}
Thus, our approximation for electrons is accurate so long as $B \sim \eta \ll 1$.

That leaves the electron neutrino, which could have a significant chemical potential if there is a large asymmetry between neutrinos and anti-neutrinos. In our universe, observations of the CMB and BBN constrain $\mu_{\nu_e} / T$ (which is conserved) to the range $[-0.01, 0.22]$ \citep{2002PhRvD..65b3511H}, which is consistent with zero but not particularly tightly constrained. Here, in common with most BBN calculations, we assume that $\mu_{\nu_e} / T$ is negligible.

In summary, our calculation gives the correct order-or-magnitude timescales for $\eta \ll 1$ (assuming no significant neutrino asymmetry). A substantial abundance of heavy elements are be produced for $\eta = 6 \times 10^{-3}$, where our assumptions should still be valid. 

A confirming set of calculations have been made by \citet{Aguirre1999,2001PhRvD..64h3508A}, who considered nucleosynthesis in a cold big bang Universe. They integrate the reaction rates, accounting for chemical potentials and degeneracy, for a reaction chain that terminates at oxygen. They note that ``by varying [the baryon-photon ratio] and [the lepton-baryon ratio], almost any desired yield of primordial helium and metals can be obtained.'' Our calculations confirm that the reactions can proceed to highly-bound elements such as iron.

A more general discussion of the connection between matter-antimatter asymmetry, BBN, fine-tuning and life can be found in \citet{Steigman2020}, who speculate that high values of $\eta$ may produce highly-bound nuclei such as iron. 

\section{Conclusions: Implications for Initial Entropy}
\label{sec:lateentropy}
The baryon-to-photon ratio in our Universe is set by the asymmetry between matter and antimatter in the early Universe. At very early times, there is roughly a billion-and-one particles for every billion antiparticles. As the Universe cools below their rest-mass energy, annihilation turns a billion particle-antiparticle pairs into two billion photons, leaving a baryon-photon ratio of $\sim$ 1 / two billion. 

Thus, an unexpected character in the story of the low entropy of our Universe emerges --- matter-antimatter asymmetry. It is well known that an exactly matter-antimatter symmetric Universe would contain essentially no baryons at late times, after annihilation. The Universe would be pure photons, unable to form structures more complex than slight overdensities of photons. If, as some have argued, black hole evaporation into photons via Hawking radiation indicates that (contra-Penrose) a pure photon gas is a higher entropy state than a black hole, then a matter-antimatter symmetric Universe would very quickly reach a state of maximum entropy, with neither nuclear nor gravitational entropy available. 

However, as we have shown here, in a universe with a sufficiently high degree of matter-antimatter \emph{asymmetry} --- which results, after annihilation, in a large value of the baryon-to-photon ratio ($\eta$) --- primordial nuclear reactions bind a substantial fraction of baryons into \Fe. Following the discussion of \citet{2019Entrp..21..466R}, in such a Universe the metastable state into which the Universe evolved would differ from its true nuclear equilibrium state at late times only with regard to a small residue of non-iron nuclei. The nuclear energy that could have been released in stars to power life has instead become a near-negligible addition to the cosmic microwave background, and is uselessly redshifting away with the expansion, unable to be put to good use.

In such a Universe, the gravitational aspect of low entropy identified by Penrose would still be available. There would still be a thermodynamic arrow of time. Matter would be able to collapse into bound structures and release energy. However, unless this collapse was able to unbind iron nuclei into its constituents, it would not ignite long-lived nuclear reaction-powered objects such as stars. 

Could collapse unbind iron? If such a process were possible, there are circumstances in our Universe where it might occur. The most extreme cases of gravitational collapse in our Universe include the accretion disks around black holes, but these show evidence of iron in their innermost regions, and no evidence of its disintegration. Type II supernovae result from the collapse of a massive star onto an iron core, but this results in nuclear reactions beyond iron (up to uranium) in the ejecta, and either a black hole or a neutron star as the relic. Again, no large scale disintegration of iron occurs. The entropic value of a neutron star seems minimal --- the nuclear energy of the neutrons does not seem to be \emph{available}.

In the absence of nuclear fuel, astrophysical sources of low-entropy radiation would have to be powered by gravitational energy. Such systems in our Universe --- protostars, accretion disks, hot stellar relics --- are shorter-lived and/or less-stable in their radiative output than stars as we know them. Note well, of course, that these scenarios may not exhaust what is possible in a pure iron Universe.

The central point is that the thermodynamic arrow of time in our Universe --- pointing from the low-entropy beginning of the Universe, through the ignition of stars by gravitational collapse, to nuclear reaction-powered low-entropy photons from the Sun, which are the source of almost all of the second-law processes around us, including ourselves --- depends crucially on matter-antimatter asymmetry in the early Universe. At the moment, we do not know what caused this asymmetry (that is, the physics of baryogenesis). We know some of the necessary conditions, thanks to \citet{1967JETPL...5...24S}. The standard model of particle physics has the required asymmetries, but not to a degree sufficient to explain our Universe. The as-yet unknown physics that explains matter-antimatter asymmetry is a major character in the story of the arrow of time.

\begin{acknowledgements}
This project was conceived at the meeting ``The Chimera of Entropy II'' held at the John Bell Center for the Foundations of Physics on the beautiful island of Hvar, Croatia in July 2019. Extreme thanks to the very useful data and code collection at \url{cococubed.asu.edu} by F.X. Timmes. This research made use of the primordial nucleosynthesis code {\it AlterBBN} ({\tt https://alterbbn.hepforge.org} \cite{2018arXiv180611095A}), the {\it SciPy} ecosystem \cite{2020SciPy-NMeth}, {\it NumPy} \cite{2020NumPy-Array} and {\it Matplotlib} \cite{4160265}. LAB acknowledges that this publication was made possible through the support of a grant from the John Templeton Foundation. The opinions expressed in this publication are those of the author and do not necessarily reflect the views of the John Templeton Foundation. We thank the anonymous referee for useful comments.
\end{acknowledgements}

\bibliographystyle{pasa-mnras}
\bibliography{main.bib}

\end{document}